# Enhancing the Authentication of Bank Cheque Signatures by Implementing Automated System Using Recurrent Neural Network

**Mukta Rao**
Kanya Gurukul Mahavidyalaya, Gurukul Kangri Vishwavidyalaya, Dehradun
Email: mukta.mca@gmail.com
**Nipur**
Kanya Gurukul Mahavidyalaya, Gurukul Kangri Vishwavidyalaya, Dehradun
Email: nipursingh@hotmail.com
**Vijaypal Singh Dhaka**
Department of Computer Science, MSIT, New Delhi
Email: vijay.dhaka@gmail.com

-------------------------------------------------ABSTRACT-------------------------------------------------
The associative memory feature of the Hopfield type recurrent neural network is used for the pattern storage and pattern authentication. This paper outlines an optimization relaxation approach for signature verification based on the Hopfield neural network (HNN) which is a recurrent network. The standard sample signature of the customer is cross matched with the one supplied on the Cheque. The difference percentage is obtained by calculating the different pixels in both the images. The network topology is built so that each pixel in the difference image is a neuron in the network. Each neuron is categorized by its states, which in turn signifies that if the particular pixel is changed. The network converges to unwavering condition based on the energy function which is derived in experiments. The Hopfield's model allows each node to take on two binary state values (changed/unchanged) for each pixel.  The performance of the proposed technique is evaluated by applying it in various binary and gray scale images. This paper contributes in finding an automated scheme for verification of authentic signature on bank Cheques. The derived energy function allows a trade-off between the influence of its neighborhood and its own criterion. This device is able to recall as well as complete partially specified inputs.  The network is trained via a storage prescription that forces stable states to correspond to (local) minima of a network "energy" function.

## 1. INTRODUCTION

Two decades ago, most banks reviewed nearly 100% of the signatures of cheque makers when they paid and manually filed cheques in file drawers in the branch. As banks grew in size and as the number of accounts, customers, and cheques processed dramatically increased, the need grew for more controlled, reliable, and efficient methods for processing and paying cheques. Accordingly, cheque storage was centralized and automated in large processing centers where the signatures of some, but not all, items were compared with the signatures on a signature card, a microfilm copy of the signature card, or a stored digital image of the signature card. With the move of the cheque pay-and-file process from the branches into large processing centers and the increase in cheque volume, signature verification changed from review of all signatures to a review of a portion of the signatures – usually large amount transactions, accounts with special arrangement to cheque multiple signatures, and a random sampling of all cheques, regardless of amount.

Handwritten signature verification has been extensively studied in this decade. Its many applications include cheques, credit card validation, security systems, certificates, contracts, etc. Comparing signature verification with fingerprint, face, voice and other recognition technologies such as retina and iris scanning, signature verification has several advantages as an identity verification mechanism:

- The capture of signature has the untouchable merits and very easy to be social accepted.
- Signature analysis can only be applied when the person is/was conscious and -disposed to write in the usual manner, although they may be under extortion to provide the handwriting [1].
- Forging a signature seems to be more difficult than a fingerprint given the availability of sophisticated analyses

However, for several reasons the task of verifying human signatures cannot be considered a trivial pattern recognition problem. It is a more difficult problem because signature samples from the same person are similar but not identical. In addition, a person's signature often changes radically during their lifetime. In fact great variability can be observed in signatures according to country, age, time, habits, psychological or mental state, physical and practical conditions [2].

There are two major methods of signature verification. One is an on-line method to measure the sequential data such as handwriting and pen pressure with a special device. The other is an off-line method that uses an optical scanner to obtain handwriting data written on paper. Handwriting recognition in off-line systems is more difficult than in on-line systems as a lot of dynamic information is lost. Hence, on -line signature verification is generally more successful. Nevertheless, off-line systems have a significant advantage in that they do not require access to special processing devices when the signatures are produced.

In fact, if the accuracy of verification could be promoted greatly, the off-line method has much more practical application areas than that of the on-line one. The process of off-line signature verification often consists of a learning stage and a testing stage. The purpose of the former is to create a reference file, and that of the latter is to compute the similarities between the testing and its corresponding reference signature to check whether the tested signature is genuine.

When we verify a signature is genuine or not, there are two types of errors: False Rejection, which is also called Type I errors; and False Acceptance, which is also called Type 2 errors. Also, there are two types of rates: False Rejection Rate (FRR) which is the percentage of genuine signatures rejected as false; and False Acceptance Rate (FAR) which is the percentage of false signatures accepted as genuine. When discussing the experiments, the results are reported in terms of FRR and FAR. But if we try to lower the FAR too much then FRR will increase. We do then have a trade off between these two types of errors and depending on the application and other aspects of where and how the system is used. We will need to optimize it in different ways.

Signature is a ballistic movement, has some characteristics such as, it is user-friendly, non invasive, as a pattern it is stored in number of applications, it is ubiquitous, it can be changed on a compromise, it is not dependent on age [3] and it is well exposed to forensic environments .The design of an accurate algorithm for verification is still an open problem [4]. Some of the different types of signatures are simple, cursive, graphical [5] and not a connected curve pattern

(Chinese signatures [6]). In off-line method, the signatures are treated as grey level images. The image can be scanned from a copy of document [7] or can be a saved file acquired by a tablet in a keyboard-less user interface.

Various methodologies have been used by researchers over the time to verify the signatures. Warping based matching methods produce the order of the coordinates and are able to match the coordinates that are placed in the same position in the two exterior curves that are formed by top and bottom parts of the template signature as test signature [8, 9].

The other effective graph matching approach is proposed by using deformable templates based on multi-resolution shape features [10]. Hidden Markov Models has ability to absorb both variability and the similarity between patterns in the problems that have an inherent temporality [11]. Forward algorithm of HMM is also used to determine the verification probability [12, 13, and 14]. Using modified direction and transition distance feature (MDF) structural features can be extracted from the signature's contour [15]. The other approaches are human centric method [16], wavelet based methods, Pseudo bacterial genetic algorithm methods(PBGA) [5] , statistical methods [17 , 18, 19] , Neural network based methods [20, 21, 22] and hybrid method [23]. Shape analysis techniques based on spatial transforms include Principal component analysis (PCA), Fisher discriminate analysis and Fast Fourier transform which extract the numerical values of features for classification.

The pseudo-dynamic features are used to detect skilled forgeries, because it is possible to recover pressure information from the image of the signature using the pseudo-dynamic features [24].

Signatures are symbols, which are usually cursively written and graphical in appearance. It is often difficult, even for humans at instances, to segment the signature images into letters or basic lexical constructs, and to recover the underlying handwriting sequence. The common approaches to off-line signature verification have been to exploit the static features of the handwriting, treating the complete signature as a single entity.

As we have discussed so far that signature verification techniques have used quite many types of Artificial Neural networks such as Multi-Layer Perceptron, Radial Basis Function Networks, Hidden Markov Models, Support Vector Machines etc. But we also have another type of very efficient and fast ANN i.e. Hopfield neural network. This network although works in unsupervised way and best suited for pattern recalling applications, but we can use its excellent feature of energy minimization for Signature verification as well.

In this paper we are proposing as network architecture for automated verification of signatures based on Hopfield neural network. The proposed network work on the idea of image change detection percentage. We enforce the network to learn the signature patterns and then recall them to compare with test sample signatures. In the following material we are presenting the quick introduction and advancements in the usage of Hopfield Neural Network.

In 1982, John Hopfield published a very influential paper which drew the attention to the associative properties of a class of neural nets [25]. The analysis was based on a definition of "energy" in the net and a proof that the net operated by minimizing this energy when settling into suitable patterns of operation [26], this helped to reignite the dormant world of Neural Networks once again. This network has attracted considerable interest both as a content address memory (CAM) and, more interestingly, as a method of solving difficult optimization problems [27-29]. The focus for solving optimization problems for researchers have been Hopfield network because of the advantages such as massive parallelism, convenient hardware implementation of the neural network architecture, fully interconnected and a common approach for solving various optimization problems [30].

Hopfield proposed two types of neural networks; Discrete Hopfield Network (DHN) and Continuous Hopfield Network(CHN). Those have been used for solving the famous travelling salesman problem in a sense of optimization [39]. DHN, a stochastic model is simple to implement and fast in computing, but it uses binary value for states of neurons resulting in an approximate solution. On the other hand, CHN gives a near-optimal solution. However, it takes too much time to simulate a differential equation which provides a main characteristic of CHN. A matching problem using a graph matching technique can be cast into an optimization problem [28].

It has been proven by Hopfield that the network converges when the weights are symmetric with zero diagonal elements (i.e. $w_{ij} = w_{ji}$ & $w_{ii} = 0$) for the hard limiting nonlinear neuron [25]. This stands true when monotonic "sigmoid" neurons with the zero diagonal elements are included in the energy function [26]. Some researchers have studied the possessions of the Hopfield neural networks with zero diagonal elements in the area of combinatorial optimization problems [31]. The experiments conducted were based on a binary Hopfield neural network with a negative self-feedback on same problem domain.  An oscillatory neuron unit was also proposed by adding a simple self-feedback, which could also be negative or positive along with an energy value extraction circuit [32-33]. However, each of them attempts to make some local minima unstable by introducing a negative self-feedback or using an oscillatory neuron.

Some researchers apply the Continuous Hopfield Network to three-dimensional object matching problem [34]. However, Continuous Hopfield Network takes much computational time in simulating a differential equation even though it provides good solutions. The Discrete type Hopfield Network has been used for two-dimensional objects matching problems [35]. However, DHN is an approximation method and gives only rough solutions, but it reduces computational time. Unlike the travelling salesman problem implemented by the Hopfield neural network [40, 41], the matching problem is handled by normalizing features made by a fuzzy function which gives distinguishable values to a connectivity matrix.

The proposed network architecture in this paper is DHN (Discrete Hopfield Network). The DHN learn the sample signature from the cheque leaves in unsupervised way, but the recalling of the same for test samples has been carried by applying constraint of threshold. The difference image is used to judge the authentication.

The rest of the paper is organized as follows. Section 2 provides a brief description of the Hopfield model as well as some terminologies to be used throughout the paper. Section 3 explains the implementation procedure. Section 4 details the pre-processing, training & test signature samples. Section 5 describes the experimental results. Finally, Section 6 concludes the paper followed by references.

## 2. TECHNICAL ASPECTS OF HOPFIELD NEURAL NETWORK

Recurrent networks are networks those have closed loops in the network topology. In this paper, a form of recurrent neural network suitable for auto associative (content) addressable memory and its characteristics, optimization and constraint satisfaction application is considered. This network is often referred to as Hopfield network in the honor of John Hopfield. The topology is very simple: it has 'n' neurons, which are all networked with each other or fully interconnected. The net has symmetric weights with no self connections i.e. all the diagonal elements of the weight matrix of Hopfield net are zero.

The two main differences between Hopfield and iterative auto associative net (recurrent associative net) are that, in the Hopfield net.

- Only one unit updates its activation at a time, and,
- Each unit continues to receive an external signal in addition to the signal from the other units in the net.

The following sections will detail all steps in the formulation and design of this proposed recurrent network and its functional details.

### 2.1 Architecture and Mathematical Foundation of Hopfield Network

The proposed system's architecture is shown in figure 1. This Hopfield network consists of a set of neurons and a corresponding set of unit delays, forming a multiple loop feedback system. A feed-back system is the one, whose output is the number of feed-back loops is equal to the number of neurons.

The HNN is a recurrent network containing feedback paths from the outputs of the nodes back into their inputs so that the response of such a network is dynamic. This means that after applying a new input, the output is calculated and fed back to modify the input. The output is then recalculated, and the process is repeated again and again. Basically, the output of each neuron is fed back, via a unit delay element, to each other neuron in the network. Successive iterations produce smaller and smaller output changes, until eventually the outputs become constant, i.e., at this moment the network achieves an acceptable stability. In this section we will first derive the Hopfield network's energy function for continuous model and by using the energy equation of continuous model we will reach to the discrete Hopfield network's energy theorem.

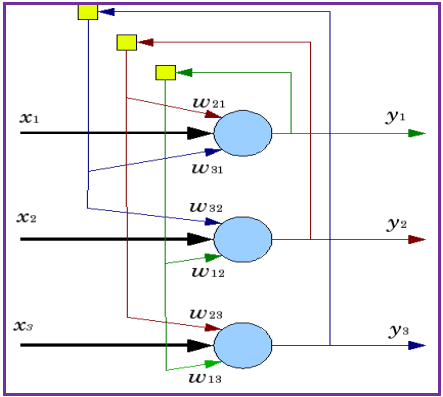

Figure 1: A Hopfield net with 3 nodes. Each node's output is fed-back to all other nodes.

The learning of signature samples is based on the energy convergence principal of Hopfield Neural Networks. The computation of HNN's energy and its dependence is discussed in this section and a pathway to energy minima is also suggested.

Hopfield network considered here is made up of neurons which possess additive model (current summing model). We know from the rule of energy convergence and voltage stability that each network follows the rules given by *Kirchoff*. From *Kirchoff's current low*, on the input node of the HNN, which has N number of neurons and at a particular time *t*, we get the equation for the HNN state as:-

$$\mathcal{C}_j \frac{dv_j(t)}{dt} = -\frac{v_j(t)}{R_j} + \sum_{i=1}^{N} w_{ji} x_i(t) + I_j,$$

For all  j = 1, 2, ....., N         (2.1)

Where

$R_j$ = leakage resistance

$x_i(t)$ = stimuli acting on synaptic weights.

$w_{ji}$ = synaptic weight.

$v_j(t)$ = induced local field at the input

$\mathcal{C}_j$ = leakage capacitance

$I_j$ = current source/ externally applied bias

Recognizing that $x_i(t) = \phi(v_j(t))$

$$\mathcal{C}_j \frac{d}{dt} v_j(t) = \frac{v_j(t)}{R_j} + \sum_{i=1}^{N} w_{ji} \varphi_i(v_i(t)) + I_j,$$

j = 1, 2, ....., N          (2.2)

Where $\phi(v_j(t))$ is the non-linear activation function.
This dynamic has got following assumptions:-
1. The matrix of synaptic weights is symmetric:

$w_{ji} = w_{ij}$ for all i and j

2. Each neuron has a *non linear activation* of its own – hence $\phi(v_j(t))$ used in equation.

3. The *inverse* of the nonlinear activation function exists, so we may write:-

$$v = \varphi_i^{-1}(x) \qquad (2.3)$$

Let the sigmoid function $\phi(v_j(t))$ be defined by the hyperbolic tangent function

$$x = \varphi_i(v) = \tanh\left(\frac{a_i v}{2}\right) = \frac{1 - \exp(-a_i v)}{1 + \exp(-a_i v)} \qquad (2.4)$$

Which has the slope of $a_i/2$ at the origin as shown by -

$$\frac{a_i}{2} = \frac{d\varphi_i}{dv}\Big|_{v=0} \qquad (2.5)$$

Henceforth we refer to $a_i$ as the gain of neuron i.

The inverse output-input relation of eq (2.4) may thus be rewritten in the form

$$v = \varphi_i^{-1}(x) = -\frac{1}{a_i}\log\left(\frac{1-x}{1+x}\right) \qquad (2.6)$$

The standard form of the inverse output-input relation for a neuron of unity gain is defined by:-

$$\varphi^{-1}(x) = -\log\left(\frac{1-x}{1+x}\right) \qquad (2.7)$$

We may rewrite eq (2.6) in terms of this standard relation as:-

$$\varphi_i^{-1}(x) = \frac{1}{a_i}\varphi^{-1}(x) \qquad (2.8)$$

The energy (Lyapunov) function $E$ of the Hopfield net work is defined by

$$E = -\frac{1}{2}\sum_{j=1}^{N}\sum_{i=1}^{N} w_{ji} x_i x_j + \sum_{j=1}^{N}\frac{1}{R_j}\int_0^{x_j}\varphi_j^{-1}(x) - \sum_{j=1}^{N} I_j x_j \qquad (2.9)$$

This energy function may have a complicated *landscape* with many minima. The dynamics of the network are described by a mechanism that seeks out those minima.

Hence differentiating $E$ with respect to time, we get

$$\frac{dE}{dt} = -\sum_{j=1}^{N}\left(\sum_{i=1}^{N} w_{ji} x_i - \frac{v_j}{R_j} + I_j\right)\frac{dx_j}{dt} \qquad (2.10)$$

By replacing values on right hand side from eq (2.2) we get

$$\frac{dE}{dt} = -\sum_{j=1}^{N} C_j\left(\frac{dv_j}{dt}\right)\frac{dx_j}{dt} \qquad (2.11)$$

We recognize the inverse relation that defined $v_i$ in terms of $x_j$. The use of eq (2.6) in eq (2.11) yields

$$\frac{dE}{dt} = -\sum_{j=1}^{N} C_j\left[\frac{d}{dt}\varphi_j^{-1}(x_j)\right]\frac{dx_j}{dt}$$

$$= -\sum_{j=1}^{N} C_j\left(\frac{dx_j}{dt}\right)^2\left[\frac{d}{dt}\varphi_j^{-1}(x_j)\right] \qquad (2.12)$$

If we graphically represent this function then we may find that inverse output-input relation $\varphi_j(x_j)$ is a monotonically increasing function of the output $x_j$. It follows therefore that

$$\frac{d}{dx_j}\varphi_j^{-1}(x_j) \ge 0 \qquad \text{for all } x_j \qquad (2.13)$$

We also note that

$$\left(\frac{d}{dt}(x_j)\right)^2 \geq 0$$
for all x_j              (2.14)

Hence, all the factors that make up the sum on the right hand side of eq 31 are non negative. In other words, for the energy function $E$ defined in Eq (2.12) we have

$$\left(\frac{d}{dt}E\right) \leq 0$$
                              (2.15)

From the definition of eq (2.15), we note that the function $E$ is bounded. Accordingly, we may make the following two statements:-

　　1) The energy function $E$ is a Lyapunov function of the Continuous Hopfield model.
　　2) The model is stable in accordance with Lyapunov's Theorem [38].

The Hopfield network may be operated in continuous mode or discrete mode, depending on the model adopted for describing the neurons. The continuous mode of operation is based on an additive model, as already discussed. On the other hand, the discrete mode of operation is based on the McCulloch-Pitts model. We may readily establish the relationship between the stable states of the Continuous Hopfield model and those of the corresponding discrete Hopfield model by redefining the input-output relation for a neuron such that we may satisfy two simplifying characteristics:-

1. The output of a neuron has the asymptotic values:-

$$X_i = \begin{cases} +1 & for \ v_i \ = \ \infty \\ -1 & for \ v_i \ = \ -\infty \end{cases}$$
                         (2.16)

2. The midpoint of the activation function of a neuron lies at the origin, as shown by:
$$\varphi_i(0) = 0$$
                              (2.17)

Correspondingly, we may set the bias Ij = 0 for all j.

In formulating the energy function E for a continuous Hopfield model, the neurons are permitted to have self loop. A discrete Hopfield model, on the other hand, need not have self loop. We may therefore simplify our discussion by setting w_ij = 0 for all j in both models.

In light of these observations, we may redefine the energy function of a continuous Hopfield model given in equation (2.9) as follows:-

$$E = -\frac{1}{2}\sum_{j=1}^{N}\sum_{i=1}^{N}w_{ji}x_i x_j + \sum_{j=1}^{N}\frac{1}{R_j}\int_0^{x_j}\varphi_j^{-1}(x)$$
                              (2.18)

The inverse function $\varphi_j^{-1}(x)$ is defined by equation (2.8).

We may thus rewrite the energy function as:

$$E = -\frac{1}{2}\sum_{j=1}^{N}\sum_{i=1}^{N}w_{ji}x_i x_j + \sum_{j=1}^{N}\frac{1}{a_j R_j}\int_0^{x_j}\varphi_j^{-1}(x)$$
                              (2.19)

The value of integral: $\int_0^{x_j}\varphi_j^{-1}(x) = 0$     (2.20)

for x_j =0, and positive otherwise. It assumes a very large value as x_j approaches ±1. If, however, the gain a_j of neuron j becomes infinitely large (i.e., the sigmoid non-linearity approaches the idealized hard-limiting form), the second term of equation no (2.19) become negligibly small. In the limiting case when a_j = ∞ for all j, the maxima and minima of the continuous Hopfield model become identical with those of corresponding discrete Hopfield model. In the case of discrete Hopfield model the energy function is defined simply by:

$$E = -\frac{1}{2}\sum_{j=1}^{N}\sum_{\substack{i=1 \\ i \neq j}}^{N}w_{ji}x_i x_j$$
                         (2.21)

Where the the j^th neuron state x_j = ±1. The equation xxx is the one that we have used in our experiments to minimize the energy for proposed system.

## 2.2  The Model as Content-Addressable Memory

John Hopfield originally designed the network as a device that would retrieve or complete certain patterns if only part of the pattern, or even a slightly distorted part of the pattern, was given to it as the starting state or input pattern. This is described as an associative memory: the network will associate or map the incomplete input with the full pattern it has 'learned' or stored, and it will retrieve this. This mapping of Hopfield network is different from Feed Forward Neural Network in the sense that it maps states into states instead of input to output. The network input is the initial state, and the mapping is through one or more states to form the network. The Hopfield networks have memory and therefore can store a set of unit outputs or system information state for use as a content-addressable memory (CAM). This network provides nearest-neighbour association of the input (initial state) with the set of stored patterns.

The content-addressable memory is such a device that returns a pattern when given a noisy or incomplete version of it. In this sense a content-addressable memory is error-correcting as it can override provided inconsistent information. The discrete Hopfield network as a memory device operates in two phases: *storage phase* and *retrieval phase*. During the storage phase the network learns the weights after presenting the training examples. The training examples for this case of automated learning are binary vectors, called also fundamental memories. The weights matrix is learned using the Hebbian Learning rule. The summation block of the *i*-th neuron is computed as

$$S_i(t+1) = f\left[\sum_{j=1}^{N}\left[w_{ji} \ S_j(t)\right]\right]$$
                         (2.22)

and next thing is to modify the weights depending on the result. There are two cases to consider:

if $s_i >= 0$ and $x_i = 0$ the output should be made negative, so each weight has to be decreased by :

$$w_{ji} = w_{ji} - (0.1 + s_i) \ / \ n$$
                         (2.23)

if $s_i < 0$ and $x_i = 1$ the output should be made positive, so each weight has to be increased by:

$$w_{ji} = w_{ij} + (0.1 - s_i) \ / \ n$$
                         (2.24)

Here learning rate coefficient is 0.1

During the retrieval phase a testing vector called probe is presented to the network, which initiates computing the neuron outputs and evolving the state. After sending the training input to the recurrent network its output changes for a number of steps until reaching a stable state. The selection of the next neuron to fire is asynchronous, while the modifications of the state are deterministic. After the state evolves to a stable configuration, that is the state is not more updated, the network produces a solution. This state solution can be envisioned as a fixed point of the dynamical network system. The solution is obtained after adaptive training.

## 2.3  Learning and Training Algorithm for the Proposed Hopfield Network

The learning in the HNN is an unsupervised process. The network learns the input sample by adjusting its weight matrix. This weight matrix change is dependent upon the output of the network which is fed back to every neuron to get the minimum state of energy. The change in weight is not an enforced constraint but it the property of HNN that lets it to achieve global energy minima. The algorithmic presentation of this learning process is explained in below mentioned algorithm 1.

| Algorithm 1: Learning in Discrete HNN |
| --- |

**Input** : Learning Signature Sample vectors in image form

**1:**  Present the given training (binary) vectors to the Hopfield net, and calculate the weights using the following method:

if $s_i >= 0$ and $x_i = 0$ : $w_{ji} = w_{ji} - ( 0.1 + s_i ) / n$    (2.25)

if $s_i < 0$  and $x_i = 1$ : $w_{ji} = w_{ji} + ( 0.1 - s_i ) / n$    (2.26)

**2.** Let the testing vector become initial state $x(0)$

**3.** *Repeat*

-*update* asynchronously the components of the state $x(t)$

$x'_i(t)$ = 1 if $S_i(t) = \sum_{j=1}^{N} \left[ w_{ji} \ x_j(t) \right] >= 0$

or $x'_i(t) = 0$ if $S_i(t) < 0$    (2.27)

-*continue* this updating until the state remains unchanged

**4.** *Until* convergence

| Output: The stable state (fixed point). |
| --- |

The execution of this algorithm results in achieving a lowest energy well for the network and the state corresponding to each element of input vector.

## 2.4  Energy Minimization

The energy is a behavioural characteristic that can be used to examine the network performance. It is well known that independently from the initial conditions the network will stabilize; it cannot oscillate even at the same energy level. The energy of Hopfield nets is defined in equation (2.21).

The energy in the Hopfield model either decreases or remains the same, the energy never increases. If it decreases it can never return to a previous state. The only case when the energy remains the same is when a neuron output changes from 0 to 1. The convergence of the network can be considered as a process of reducing the network energy until the network does not reach to the minimum energy state.

## 3. IMPLEMENTATION PROCEDURE

As already briefed, the method of signature verification proposed in this paper, is based on pixel based difference and allowed percentage limit of change in signature. So each element or in other terms each pixel contributes in the determination of the authenticity of the cheque.

Under the analogy HNN paradigm, the problem of signature verification is to label each pixel of the incoming images as changed or unchanged and the degree of change. This change is computed by recalling the pattern stored by HNN and also by getting the difference image matrix. Both of the truth matrixes are compared to decide the percentage of change in both the images of signature.

With such purpose, we consider the output image as a network of nodes where each node is associated to a pixel location in the difference image, i.e., the number of nodes is exactly the number of pixels of the incoming images. Also, each node is characterized by its state value vi, ranging in [-1, +1]. The network state is characterized by the states of the nodes. After the optimization process, when the network stability is reached, the nodes have achieved its final state value.

This final value will indicate unchanged (-1) or maximum change (+1); intermediate values give the strength of the change. The optimization process minimizes the energy in (2.21) starting from an initial network state. The interconnection weights in (2.25) are computed by applying data and contextual consistencies and the external inputs through the self-data information. The procedure can be depicted using following diagram in figure 2.

Training a Hopfield net involves lowering the energy of states that the net should "remember". This allows the net to serve as a content addressable memory system, that is to say, the network will converge to a "remembered" state if it is given only part of the state. The net can be used to recover from a distorted input the trained state that is most similar to that input. This is called associative memory because it recovers memories on the basis of similarity. Thus, the network is properly trained when the energy of states which the network should remember are local minima.

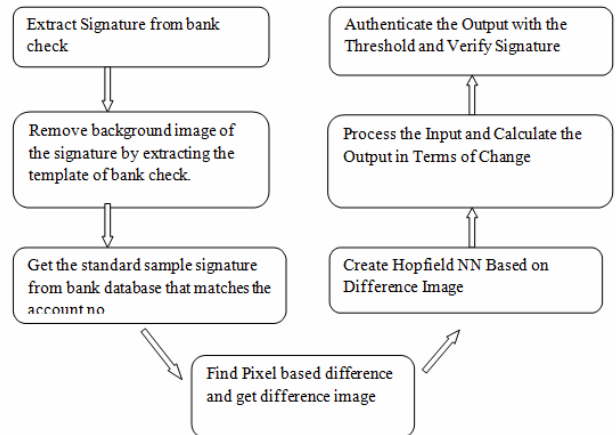

Figure 2: Overall processing of proposed Signature verification system based on HNN.

Each step outlined in the flow diagram (figure 2) is discussed in detail in sections 3.1 to section 3.4.

## 3.1  Pre-Processing and Network Initialization

The signature on the bank Cheque is always containing the back-ground, because of the reason that the Cheque leaves cannot be forged. Although this is a security feature for the bank but it leads to specific module in our system. We have to pre-process the Cheque's image before it can be presented to the system.

The back-ground image of a signed bank Cheque's leaf can be removed in two ways:

- It is evident from the real-life samples that the back-ground image's intensity is always very less as compared to the signature's ink density. Hence the image of signature can be binarized in way that the intensity threshold removes the back-ground.

- In this method we can subtract the standard template back-ground image from the one that is signed signature, this subtraction should be done as a Homomorphic image subtraction (gray scale based using equation 2.28).

For the proposed system the input image is changed based on above described method 2. The network initialization is carried out by exploiting the characteristics of the difference image. We use the initialization strategy, described in [36, 37], as follows. From the histogram of the difference image, we compute thresholds. Now, given a pixel location in the difference image, its associated node in the network is initialized based on this threshold. In the work of Wang [38], a list of major grouping principles is exhaustively studied.

In our signature verification approach, we apply the following two principles:

- Proximity
- Changed/unchanged pixels that lie close in space tend to group.

The State of each pixel is computed by HNN and its confirmation is done as based on following computations. First of all the difference image of both of the signatures (sample and test) is computed. The Difference Image is computed as:-

$$D(x,y) = I_1(x,y) - I_2(x,y) \qquad (2.28)$$

Here $I_1$ and $I_2$ are the binary Matrixes of the Learned and test signatures. The two dimensional indices are used in terms of x and y.

Initial State of HNN is determined as =

$O(x,y) = 1$, if $D(x,y) >$ Threshold (changed)

$\quad = 0$, Otherwise   (no change) $\qquad (2.29)$

## 4. TRAINING & TEST SIGNATURE SAMPLES AND EXPERIMENTS

The success of any technique is closely dependent on the samples on which the experiments have been conducted. The samples of signatures were gathered from academic staff of University and Government school. The members gave signature samples on blank sheet which was cross marked but signed Cheque leaves were provided for verification purpose as well. 83 such samples were gathered and processed (see appendix 1 for samples). 40 samples were consumed during learning process of HNN where as rest of the 40 samples were used for testing of the system. Three samples were discarded because of being distorted. During the experiments one intermediate image was obtained using equation 2.28. That is a truth image; which is the simple difference image that state if the pixel is changed or not, this has been derived as a binary image. As shown in the following figures:

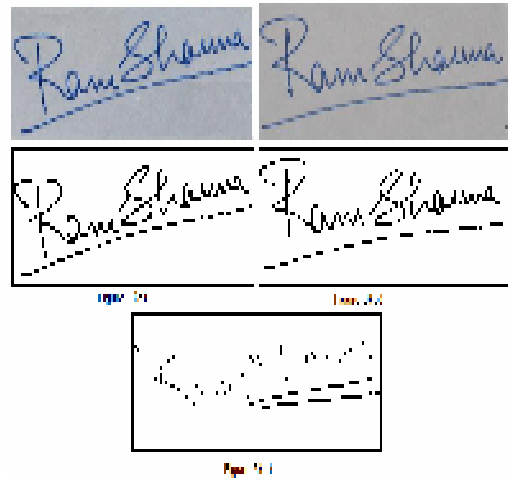

Figure 3(a): Sample signature of the customer stored in the Bank's DataBase (original and binarized). Figure 3(b): Signature of the customer captured from the Cheque (original and binarized). Figure 3(c): Difference Image calculated from the difference of both binary images of the signature.

To deduce the network performance the percentage of correctly labelled pixels was computed. Correct labelling means to decide the current truth table to each of the pixel. Any pixel of the image can be labelled as one state value out of the following four:-

1. The pixel is not at all changed in both images but the HNN wrongly computes as changed.
2. The Pixel is changed and the HNN wrongly determines as unchanged.
3. The pixel is not changed in both images and HNN also states that it is not changed.
4. The pixel is changed and HNN correctly determines as changed.

So based on the knowledge of the state of each pixel we can compute the percentage of success of HNN in determination of the state, using following formula:-

$$Perc = ((NT + PT)/(NT+PT+NF+PF))*100 \qquad (2.30)$$

Where

NT = Negative True; i.e. number of no change pixels correctly detected by HNN

PT = Positive True; i.e. number of changed pixels correctly detected by HNN

NF = Negative False; i.e. number of no change pixels incorrectly detected by HNN

PF = Positive False; i.e. number of changed change pixels not detected by HNN

In other terms we can say that NT belongs to state category 3, PT belongs to state category 4, NF belongs to state category 2 and PF belongs to state category 1 from the possible state categories as earlier mentioned for each pixel. Detecting by the HNN simply means as labelling of the state value. The judgement of the signature verification was based on the higher value of the Perc. The network was made o continue the iterations until it reached to the 95% of percentage (network goal). When done so, the number of actual changed pixels were calculated and compared to the threshold. If the threshold is lower that this number, signature is said to be forged.

## 5. EXPERIMENTAL RESULTS AND DISCUSSION

The performance of the system is analysed in terms of the correct verifications, computation time and number of iteration cycles (network iterations). The table 1 shows the number of correctly verified signatures which were actually correct and incorrect and also the discarded or failed signatures which were actually correct and incorrect. This way we achieve 4 values in total. These values are presented as percentage in table 1.

| Type | Verified | Discarded |
|------|----------|-----------|
| Genuine Signatures | 82% | 18% |
| Forge Signatures | 11% | 89% |

Table 1: Signature Verification Data Result Statistics

In this table column 1 signifies the percentage of passed signatures. And Column 2 represents failed/discarded signatures. The two rows represent the type of signatures. Genuine signature means both learnt and test sample signatures are captured from the same person, so they all should be passed, but as we can see only 82 percent signatures could be verified correctly. 18 percent signatures were discarded even though they were not forged.

On the other hand we can see that in table 1 above the second row represent the signature samples which are forged at the time of testing, so the learnt signature sample and testing signature samples were gathered from different people. So ideally the system should have failed all the signatures, but only 89 percent accuracy could be obtained.

The network's learning and testing (as such recalling and threshold based verification) has been carried out in multiple sets and during each set of execution, we have taken Energy state and iterations into account. The following table (table no 2) represents the comparison between the learning and test cycle in terms of iteration cycles.

| Iteration cycles | Training Data Set Energy (E x $10^6$) | Testing Data Set Energy (E x $10^6$) |
|---|---|---|
| 1 | -1.1167 | -0.7167 |
| 2 | -1.893 | -1.193 |
| 3 | -2.4 | -1.26 |
| 4 | -2.7 | -2.1 |
| 5 | -3.3 | -2.43 |
| 6 | -4.1 | -2.91 |
| 7 | -4.5 | -3.25 |
| 8 | -4.7 | -3.3 |
| 9 | -4.8 | -3.6 |
| 10 | -4.84 | -3.65 |

Table 2: Hopfield Neural Network energy values for the training and test data sets.

The column 1 of table 2 represents the iteration number at which the energy values of both of the networks have been observed. Each of the readings mentioned in table 2 is a simple average of energy values computed for all training and test samples. Column 2 represents the energy values computed in the case of training data and column 3 represents energy values in the case of test pattern signatures.

The graphical representation of data from above mentioned table 2 is shown below in Figure 4. This figure shows the average of energy values for verification of correct as well as incorrect training and testing data sets.

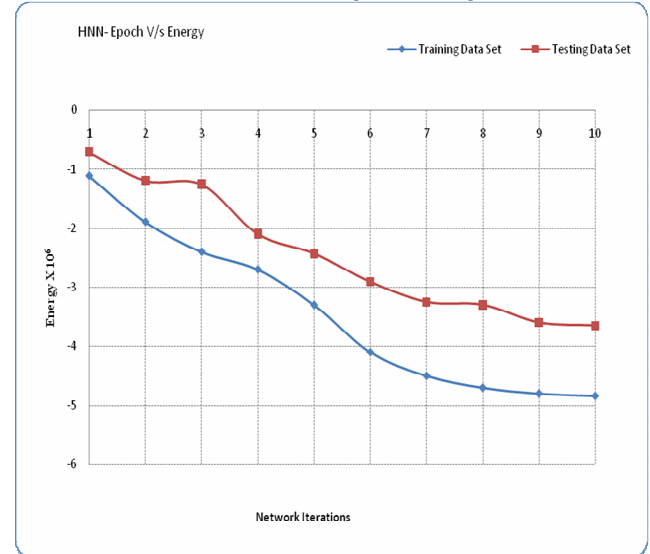

Figure 4: Comparison of average energy value convergence of the system with training and test samples of the signature.

Figure 4 depicts that the energy minimization in the case of training and test patterns are compared. The proposed system achieves less energy values in relatively less time and efforts in the case of training data, whereas; the test patterns consumed more time and efforts to converge. The gradual and almost constant rate of energy minimization of Hopfield neural network is observant from figure 4.

As we have discussed in section 3 that for the network's best performance and for the judgement of test signature (in terms of forged/ passed), we compute the percentage of the correct labelling done by HNN. This percentage is also a key role player in network's iterations. We have captured the percentage values for the network for both learning and testing samples. An average percentage is shown in table 3.

| Iteration cycles | Training Data Set Percentage | Testing Data Set Percentage |
|---|---|---|
| 1 | 0.46 | 0.54 |
| 2 | 0.51 | 0.57 |
| 3 | 0.57 | 0.58 |
| 4 | 0.6 | 0.63 |
| 5 | 0.67 | 0.64 |
| 6 | 0.76 | 0.67 |
| 7 | 0.82 | 0.7 |
| 8 | 0.86 | 0.72 |
| 9 | 0.88 | 0.745 |
| 10 | 0.9 | 0.75 |

Table 3: HNN correct state percentage values for the training and test data set.

The percentage values presented in table 3 are decided based on the two way difference images. Column 1 contains the behaviour of Hopfield neural network in the case of training samples and column 3 represents percentage data of test signatures. The graphical representation of data from Table 3 is shown in figure 5.

It is evident from table 3 that network takes more iterations to correctly label the test samples; one possible reason could be that we have mixed test data set with forged and genuine signatures. Due to presence of in-correct data in test data set the HNN is taking more time to correctly label all pixels / elements in test vector.

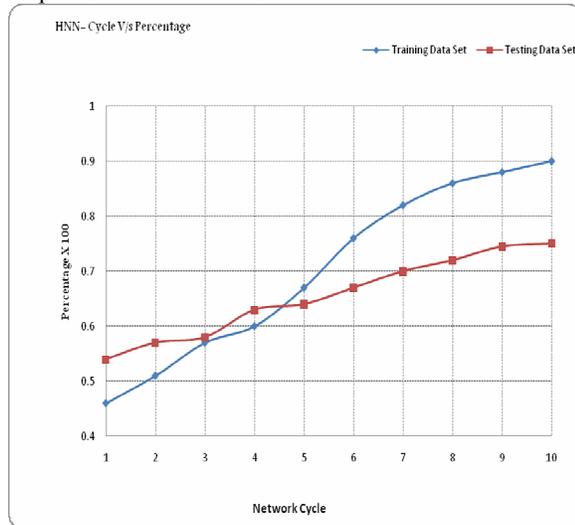

Figure 5: The Percentage values computed from equation (2.22) against the network iterations for training and test data sets.

## 6. CONCLUSION

In this paper, we have developed a new automatic strategy for verification of authentic signatures in bank Cheques using Hopfield Neural Networks. Some researchers have developed strategies on pattern recognition using HNN; we have extended their work for the specific area of signature verification. We have also used this kind of information for integrating the data and contextual information in form of an energy function that converges rapidly. The mapping of data information is improved in relation to classical implementations of this information. This improvement is achieved through the inter-data relations.

As it is visible from Figure 4, the convergence is fast and the results of the HNN are better in the sequence without significant illumination changes. The results in Table 2 display that the performance of network is remarkable. The best performance in terms of accuracy is obtained with iterative methods. Part of the improvement of HNN is due to the mapping of the self-information applied by HNN. This takes special relevance near the borders of changed areas. The initialization process can be important but it is not decisive. The noise removal process of the Cheque's image is still a big challenge for this strategy.

The main drawback of Hopfield network is the high evaluation time, so for real life situations and faster responses, this strategy should be implemented on parallel architectures. As we can see from the first row of Table 1 that although, all signatures were genuine and drawn by the same person who gave learning sample; but still the system discarded 18 percent of signature samples. Similarly from the second row of table 1 we can see that the proposed system is unable to discard 11 percent of the forged signatures. So this is the shortcoming of the proposed system and this can be further enhanced. More work needs to be carried out in this domain to get higher accuracy and great speed.

## Authors Biography

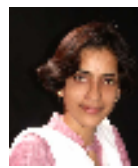
*Mukta Rao* is currently pursuing his Ph.D in computer science from Gurukul Kangri Vishwavidyalaya and currently working with Interglobe Technologies Ltd,

Gurgaon, Haryana, India as a senior software engineer. She has done her MCA from University of Rajasthan, India. She has been associated with Hughes Software Systems and Infosys. Her research areas include Handwriting recognition, Neuro-computing, Web-Acceleration and Network security.

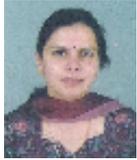

*Nipur* received a PhD from Gurukul Kangri Vishwavidyalaya and an MCA from Banasthali Vidyapith. Presently she is Reader, Department of Computer Science, Kanya Gurukul Mahavidyalaya, Dehradun (second campus of Gurukul Kangri Vishwavidyalaya,Hardwar). She has 15 years PG teaching experience. Her recent research has been in the field of mobile computing & communication and interconnection networks. She handled many research projects during last ten years; Downlink Scrambling code grouping scheme for UTRA, modified generator for multiple scrambling codes, comparison of cell search methods for CDMA systems, etc She is the author of more than 35 research papers in international, national journals and conference proceedings. She is a life member of CSI & SSI and member of IEEE & IEEE Computer Society.

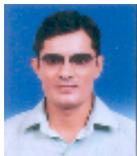

*Vijaypal Singh Dhaka* is received his Ph.D. in computer science from Dr.  BR Ambedkar University, Agra, UP, India. He is currently a Sr.Lecturer  in department of computer science and engineering ,Maharaja Surajmal Institute of Technology, New Delhi, India. He completed his M. Phil. in computer science in 2007. His research interests are neuro-computing, character recognition, soft-computing and internet technologies.